\documentclass[%
 reprint,
 superscriptaddress,
 bibnotes,
 amsmath,amssymb,
 aps,
 prl,
]{revtex4-2}

\usepackage{graphicx}
\usepackage{dcolumn}
\usepackage{bm}

\usepackage[disable]{todonotes}
\usepackage[version=4]{mhchem}

\newcommand{\abs}[1]{\left\lvert#1\right\rvert}
\newcommand{\dd}{\,d}               

\newcommand{\kT}{k_{\mathrm{B}}T}   
\newcommand{\x}{\bm{x}}             
\newcommand{\y}{\bm{y}}             
\newcommand{\s}{\bm{s}}             
\newcommand{\map}{\mathcal{M}}      
\newcommand{\maps}{\map_{\s}}       
\newcommand{\Jm}[1]{\abs{J_{\map}(#1)}} 
\newcommand{\logJm}[1]{\log\Jm{#1}} 
\newcommand{\deltas}[1]{\delta\left[ \s(#1){-}\s \right]}   
\newcommand{\btheta}{\bm{\theta}}   
\newcommand{\bthetaoptim}{\btheta^*}   
\newcommand{\bthetaml}{\btheta^{\mathrm{ML}}}   
\newcommand{\train}{\mathrm{tr}}     
\newcommand{\eval}{\mathrm{ev}}      

\begin{document}

\title{Targeted free energy perturbation revisited: \\ Accurate free energies from mapped reference potentials}

\author{Andrea Rizzi}
 \affiliation{Computational Biomedicine, Institute of Advanced Simulations IAS-5/Institute for Neuroscience and Medicine INM-9, Forschungszentrum J{\"u}lich GmbH, J{\"u}lich 52428, Germany}
 \affiliation{Atomistic Simulations, Italian Institute of Technology, Via Morego 30, Genova 16163, Italy}
\author{Paolo Carloni}
 \email{p.carloni@fz-juelich.de}
 \affiliation{Computational Biomedicine, Institute of Advanced Simulations IAS-5/Institute for Neuroscience and Medicine INM-9, Forschungszentrum J{\"u}lich GmbH, J{\"u}lich 52428, Germany}
 \affiliation{Molecular Neuroscience and Neuroimaging (INM-11), Forschungszentrum J{\"u}lich GmbH, J{\"u}lich 52428, Germany}
 \affiliation{Department of Physics and Universit{\"a}tsklinikum, RWTH Aachen University, Aachen 52074, Germany}
\author{Michele Parrinello}
 \email{michele.parrinello@iit.it}
 \affiliation{Atomistic Simulations, Italian Institute of Technology, Via Morego 30, Genova 16163, Italy}


\begin{abstract}
We present an approach that extends the theory of targeted free energy perturbation (TFEP) to calculate free energy differences and free energy surfaces at an accurate quantum mechanical level of theory from a cheaper reference potential.
The convergence is accelerated by a mapping function that increases the overlap between the target and the reference distributions.
Building on recent work, we show that this map can be learned with a normalizing flow neural network, without requiring simulations with the expensive target potential but only a small number of single-point calculations, and, crucially, avoiding the systematic error that was found previously.
We validate the method by numerically evaluating the free energy difference in a system with a double-well potential and by describing the free energy landscape of a simple chemical reaction in the gas phase.
\end{abstract}

\maketitle


\textit{Introduction.}---Predicting free energy changes is one of the fundamental problems in physics and chemistry with countless applications to drug development, biology, and materials science~\cite{chipot2007free}.
Molecular simulations provide a rigorous means of determining this property.
Endpoint and alchemical approaches~\cite{wang2019end,mey2020best} calculate free energy differences (FED) between two states (e.g., binding and solvation free energies) or two different molecules (e.g., the change in affinity between two drugs towards their target receptor).
Instead, methods such as umbrella sampling~\cite{torrie1977nonphysical,souaille2001extension} (US), metadynamics~\cite{laio2002escaping,barducci2008well,invernizzi2020rethinking} (MetaD), and adaptive biasing force~\cite{darve2001calculating,henin2004overcoming} allow reconstructing the free energy surface (FES) of a system as a function of one or more physical collective variables (CVs).

While much progress has been made in developing empirical potentials capabable of accurate predictions~\cite{gapsys2020large,kuhn2020assessment,lee2020alchemical}, a full or partial quantum mechanical (QM) representation of the system was often found to be desirable.
For instance, hybrid molecular mechanics/quantum mechanics potential energy surfaces based on post-Hartree-Fock, density functional theory (DFT), or quantum machine learning potentials~\cite{behler2007generalized,schutt2018schnet,smith2019approaching,noe2020machine} have been used to overcome some of the shortcomings of empirical and semi-empirical calculations to study ligand binding~\cite{wang2019host,dybeck2016comparison,hudson2018force,capelli2020accuracy,rufa2020towards} and chemical reactions \cite{piccini2019accurate,sirirak2020benchmarking,pan2019accelerated,shen2018molecular}.
However, the use of accurate quantum chemical methods severely limits the size of the system that can be studied.

In this respect, using a cheaper potential as a reference and recovering the accuracy of a more expensive Hamiltonian with free energy perturbation (FEP)~\cite{zwanzig1954high} may achieve significant computational savings.
Indeed, only a relatively small number of expensive energy calculations are required to achieve the accuracy of the target level of theory.
This approach was pioneered by Gao~\cite{gao1992absolute} and by Muller and Warshel~\cite{muller1995ab} for the calculations of the FED and FES, respectively.
However, the speed of convergence of FEP degrades rapidly as the overlap between the reference and target distributions decreases, restricting its applicability~\cite{olsson2016converging,pan2019accelerated}.

To overcome this problem, several methods have been successfully developed so as to avoid extensive simulations of the full system with the target Hamiltonian.
These either use a sequence of intermediate Hamiltonians~\cite{olsson2017comparison,wang2019host}, or employ highly efficient estimators~\cite{dybeck2016comparison,li2018accelerated}, or train cheaper and \textit{ad-hoc} parametrized models for either the reference~\cite{hudson2018force,shen2018molecular,pan2019accelerated} or the target potential~\cite{shen2016multiscale,chehaibou2019computing,bucko2020ab}.
An elegant approach to the calculation of the FED was introduced by Jarzynski in 2002 under the name of targeted free energy perturbation (TFEP)~\cite{jarzynski2002targeted,hahn2009using}.
The method performs a mapping of the atomic coordinates such that the overlap between reference and target distributions is increased.
An advantage of TFEP over existing methods is that, in principle, it enables instantaneous convergence of the free energy difference.
Moreover, the mapped configurations can also be used to study the molecular geometries at the target level of theory.
Because this mapping is generally very complex, it has been suggested to represent it with a neural network (NN)~\cite{wirnsberger2020targeted}.
Unfortunately, such NN was found to introduce a systematic error unless it was also trained with samples obtained from extensive simulations using the target Hamiltonian.
In the context of reference potential methods, however, this defeats the purpose of employing a cheaper Hamiltonian.

In this Letter, we revisit TFEP and the formulation of the learning problem proposed in Ref.~\cite{wirnsberger2020targeted}, we find the origin of the difficulty, derive a rigorous bound that explains the systematic error reported and, above all, offer a solution.
As a result, we are able to learn an efficient Jarzynski mapping with a small set of QM energy and gradient calculations.
Next, we validate the methodology numerically on a simple, double-well potential system.
Finally, we extend the method to the calculation of the FES and test it by simulating a simple chemical reaction in the gas phase.

\textit{Computing free energy differences.}---Consider the problem of determining the free energy difference between a reference (or sampled) distribution $A$ and a target distribution $B$, which is given by
\begin{equation}
    \Delta f_{AB} = - \log\frac{Z_B}{Z_A} = -\log\frac{\int_{\Gamma_B} e^{-u_B(\y)} \dd\y}{\int_{\Gamma_A} e^{-u_A(\x)} \dd\x} \; ,
\label{eq:delta-f_AB}
\end{equation}
where for convenience we have expressed free and potential energies in units of $\kT$, and $Z_A$, $\Gamma_A$, and $u_A(\x)$ are the configurational partition function, the domain of integration, and the reduced potential energy, respectively, of configuration $\x$ associated with the Boltzmann distribution $A$
\begin{equation}
    p_A(\x) = \frac{e^{-u_A(\x)}}{Z_A} \; .
\label{eq:boltzmann-distribution}
\end{equation}
When $\Gamma_A = \Gamma_B$, Eq.~(\ref{eq:delta-f_AB}) can be computed using only samples from $A$ through the Zwanzig identity~\cite{zwanzig1954high}
\begin{equation}
    \Delta f_{AB} = -\log \langle e^{-w_{AB}(\x)} \rangle_A \; ,
\label{eq:fep}
\end{equation}
where $\langle g(\x) \rangle_A = \int_{\Gamma_A} p_A(\x) g(\x) \dd\x$ and $w_{AB}(\x) = u_B(\x) - u_A(\x)$ may be interpreted as the work performed in an infinitely fast non-equilibrium process switching the system from $A$ to $B$~\cite{jarzynski1997nonequilibrium}.

Jarzynski's TFEP derivation~\cite{jarzynski2002targeted} is based on an invertible transformation $\map: \Gamma_A \to \Gamma_B$ that maps samples from $A$ to a different ensemble $A'$ that shares a larger overlap with $B$.
We provide here an alternative derivation and interpretation of TFEP that considers the map $\map$ as transforming the target distribution $B$ rather than $A$.
Specifically, we exploit the work of Zhu \textit{et al.}~\cite{zhu2002using}, who noticed that, under a change of variable $\y = \map(\x)$, the configurational partition function becomes
\begin{equation}
    Z_B = \int_{\Gamma_A} e^{- u_B(\map(\x)) + \logJm{\x}} \dd\x \; ,
\label{eq:partition-function-bprime}
\end{equation}
where $|J_{\map}|$ is the absolute value of the Jacobian determinant of $\map$.
The key observation is the appearance of an effective potential
\begin{equation}
    u_{B'}(\x|\map) = u_B(\map(\x)) - \logJm{\x}
\label{eq:potential-bprime}
\end{equation}
which defines a new Boltzmann distribution $B'$ on $\Gamma_A$.
Importantly, $\map$ transforms $B$ by reshaping its potential but without changing $Z_B$ (and thus its free energy).
The problem in Eq.~\ref{eq:delta-f_AB} is then equivalent to computing $\Delta f_{AB'} = -\log Z_{B'}/Z_A$ for a convenient choice of $B'$, and we can recover the TFEP estimator~\cite{jarzynski2002targeted} by simply applying FEP on $A$ and $B'$
\begin{equation}
    \Delta \hat{f}_{AB}(D|\map) = -\log \frac{1}{N} \sum_i^N e^{-w_{AB'}(\x_i|\map)} \; ,
\label{eq:tfep-estimator}
\end{equation}
where $D = \{\x_i\}$ is a dataset of $N$ samples from $p_A(\x)$, and $w_{AB'}(\x|\map) = u_{B'}(\x|\map) - u_A(\x)$.
The estimator in Eq.~(\ref{eq:tfep-estimator}) converges to $\Delta f_{AB}$ for any invertible map, but its convergence rate strongly depends on the choice of $\map$.
The optimal choice $\map^*$ transforms $B$ so that $B'$ overlaps perfectly with $A$, i.e.,
\begin{equation}
    p_{B'}(\x|\map^*) = \frac{e^{-u_{B'}(\x|\map^*)}}{Z_B} = p_A(\x) \; .
\label{eq:optimal-map-defining-condition}
\end{equation}
Using Eq.~(\ref{eq:boltzmann-distribution}) and (\ref{eq:optimal-map-defining-condition}), it is easy to obtain~\cite{jarzynski2002targeted}
\begin{equation}
    w_{AB'}(\x|\map^*) = \Delta f_{AB} \; .
\label{eq:optimal-work-tfep}
\end{equation}
A remarkable feature of this relation is that, with $\map^*$ at hand, a single sample is sufficient to obtain a converged estimate of $\Delta f_{AB}$.

Following Ref.~\cite{wirnsberger2020targeted}, we implement the map with a suitably trained normalizing flow neural network~\cite{papamakarios2021normalizing}.
Normalizing flows are particularly suited for this problem since they lead by construction to maps that are invertible and whose Jacobian (see Eq.~(\ref{eq:potential-bprime})) can be computed cheaply.
The parameters $\bthetaoptim$ of the NN representing the optimal map $\map^*$ are obtained by minimizing the Kullback-Leibler (KL) divergence $D_{\mathrm{KL}}\left[ p_A || p_{B'} \right]$, which can be written as
\begin{equation}
\begin{split}
    D_{\mathrm{KL}}\left[ p_A || p_{B'} \right] &= \int_{\Gamma_A} p_A(\x) \log \frac{p_A(\x)}{p_{B'}(\x|\btheta)} \dd\x \\
    &= \langle w_{AB'}(\x | \btheta) \rangle_A - \Delta f_{AB} \; .
\label{eq:tfep-kl-divergence}
\end{split}
\end{equation}
In Ref.~\cite{wirnsberger2020targeted}, $\Delta \hat{f}_{AB}$ is then computed with the optimized map on the same dataset used to train the NN.
This choice was motivated by the goal of reducing the amount of data needed.

Here, we show that such a procedure leads to a systematic error.
To this end, let us first recast the learning problem as a maximum likelihood (ML) estimation.
Given a set $D_{\train}$ of $N_{\train}$ independent samples from $p_A$, we use Eq.~(\ref{eq:optimal-map-defining-condition}) to write the probability of observing the data as
\begin{equation}
    p_A(D_{\train}) = \prod_i^{N_{\train}} p_{B'}(\x_i|\map^*) \propto \prod_i^{N_{\train}} e^{-u_{B'}(\x_i|\map^*)} \; .
\end{equation}
After multiplying by $\prod_i e^{u_A(\x_i)}$, which do not depend on the map, and simple manipulation, the negative log-likelihood reads:
\begin{equation}
    \mathcal{L}(\btheta) = \frac{1}{N_{\train}} \sum_i^{N_{\train}} w_{AB'}(x_i | \btheta) \; .
\label{eq:loss-tfep}
\end{equation}
The expression in Eq.~(\ref{eq:loss-tfep}) is, within an immaterial constant, how the KL divergence in Eq.~(\ref{eq:tfep-kl-divergence}) is estimated using a finite sample.
In the limit of $N_{\train} \to \infty$, $\mathcal{L}(\btheta)$ approaches the true value of the KL divergence, and the ML solution $\bthetaml$ is identical to the perfect map $\bthetaoptim$.
In general, however, with a finite sample we have by definition $\mathcal{L}(\bthetaml) \le \mathcal{L}(\bthetaoptim)$.
Thus, using Eq.~(\ref{eq:optimal-work-tfep}) and Jensen's inequality we finally obtain the inequality
\begin{equation}
    \Delta \hat{f}_{AB}\left( D_{\train} | \bthetaml \right) \le \Delta f_{AB} \; .
\label{eq:train-df-bound}
\end{equation}
This is the cause for the systematic error observed in Ref.~\cite{wirnsberger2020targeted}.
Eq.~(\ref{eq:train-df-bound}) predicts that if the free energy estimate is computed on the training dataset, it converges systematically to an incorrect value as the training increases.
Such error can thus be interpreted as a peculiar case of overfitting that results in performance degradation on the training set.

To solve this problem, instead of using samples from the target distribution as in Ref.~\cite{wirnsberger2020targeted}, we note that Eq.~(\ref{eq:train-df-bound}) is a consequence of the dependence of $\bthetaml$ on $D_{\train}$.
Thus, the issue vanishes if $\Delta \hat{f}_{AB}$ is computed using an \textit{independent} set of configurations.
Indeed, given such an evaluation dataset $D_{\eval}$ of $N_{\eval}$ samples from $p_A$, by applying Jensen's inequality to Eq.~(\ref{eq:tfep-estimator}) we recover the bound~\cite{hahn2009using}
\begin{equation}
    \langle \Delta \hat{f}_{AB}(D_{\eval} | \btheta) \rangle \ge \Delta f_{AB} \; ,
\label{eq:eval-df-bound}
\end{equation}
where the mean is intended over all possible instances of $D_{\eval}$.
Eq.~(\ref{eq:eval-df-bound}) is valid for any map $\btheta$, including $\bthetaml$, and the equality is approached for $N_{\eval} \to \infty$.

Note that a systematic error is present also when using the evaluation set but only on average rather than on \textit{any} training datasets as in Eq.~(\ref{eq:train-df-bound}).
Moreover, we expect the error on $D_{\train}$ to increase in the low-data regimes typical of molecular simulations, where only a few thousand independent data points can be used for training, and the effects of overfitting are more pronounced.
Thus, in practice, we expect to obtain smaller errors and more robust estimates of the uncertainty of $\Delta \hat{f}_{AB}$ on an independent set than on $D_{\train}$.

\begin{figure}[t]
    \includegraphics{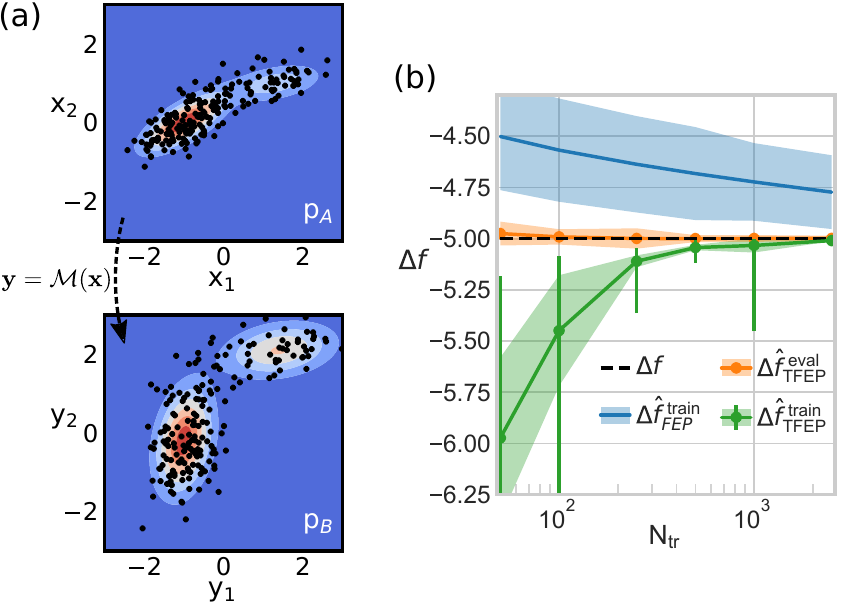}
    \caption{\label{fig:toy-problem}Calculation of $\Delta {f}_{AB}$ on a 2D free energy landscape.
    (a) Contour plots of the sampled (A, top) and target (B, bottom) distributions.
    The overlaid black dots were sampled from $p_A$ (top) and mapped through the trained $\map$ (bottom).
    (b) Theoretical free energy difference (dashed, black) or computed on the evaluation dataset ($N_{\eval} = 100$) with TFEP (orange) and on the training set with TFEP (green) or FEP (blue).
    Lines, shaded areas, and (for $\Delta \hat{f}^{\mathrm{train}}_{\mathrm{TFEP}}$) error bars represent the average, standard deviation, and the full range of the predictions across repeats, respectively.
    The energy values are in units of $\kT$.}
\end{figure}

We check the validity of our method by computing $\Delta {f}_{AB}$ between the two double-well potential distributions in Fig.~\ref{fig:toy-problem}.
The parameters of the two distributions were chosen to be sufficiently different for standard FEP to exhibit slow convergence (Fig.~\ref{fig:toy-problem}(b)).
We implemented the parametric map with an inverse autoregressive flow (IAF)~\cite{kingma2016improved,germain2015made}.
For each tested training dataset size $N_{\train}$, we trained 40 NNs on randomly-generated datasets and evaluated each network on independent datasets of constant size $N_{\eval} = 100$.
We refer the reader to the Supplemental Material~\cite{supplementalmaterial} for the technical details of all the numerical experiments in this work.

As predicted by Eq.~(\ref{eq:train-df-bound}), every single NN underestimated $\Delta f_{AB}$ on $D_{\train}$, and, counterintuitively, overfitting resulted in significantly worse performance on the training rather than the independent set, especially when $D_{\train}$ was small.
In contrast, in spite of its small size ($N_{\eval} = 100$), evaluating $\Delta \hat{f}_{AB}$ on the evaluation sets consistently outperformed both standard FEP and TFEP on $D_{\train}$.
The data in Fig.~\ref{fig:toy-problem} support the conclusion that it is beneficial to reduce the amount of data available for training to allow the calculation of $\Delta \hat{f}_{AB}$ on a set of independent configurations.

\textit{Computing free energy surfaces.}---Here, we show how TFEP theory can be extended to determine the free energy surface $f_B(\s)$ as a function of a generally multidimensional CV $\s$.
We do so by adding to $f_A(\s)$, which is estimated using a cheap reference potential, the perturbation term
\begin{equation}
    \Delta f_{AB}(\s) = -\log\frac{Z_B(\s)}{Z_A(\s)} = -\log \langle e^{-w_{AB}(\x)} \rangle_{A|\s} \; ,
\label{eq:dfs-identity}
\end{equation}
where $Z_B(\s) = \int_{\Gamma_B} e^{-u_B(\y)} \deltas{\y} d\y$ is the partition function restricted to $\s$, and the average is taken over the distribution
\begin{equation}
    p_A(\x|\s) = \frac{e^{-u_A(\x)}}{Z_A(\s)}\deltas{\x} \; .
\end{equation}
First, note that simply plugging in Eq.~(\ref{eq:dfs-identity}) the optimal TFEP map $\map^*$, using Eq.~(\ref{eq:optimal-work-tfep}), results in the incorrect expression $\Delta f_{AB}(\s) = \Delta f_{AB}$.
Indeed, to formulate an equivalent problem in terms of the transformed distribution $B'$, the change of variable must now preserve the value of $Z_B(\s)$.
To achieve this, a sufficient condition is that the transformation $\maps: \Gamma_A \to \Gamma_B$ satisfies
\begin{equation}
    \s\left(\maps(\x)\right) = \s(\x) \; .
\label{eq:cv-preserving-condition}
\end{equation}
The relation in Eq.~(\ref{eq:cv-preserving-condition}) prevents $\maps$ from moving probability density along $\s$ so that $p_B(\s) = Z_B(\s)/Z_B$ (i.e., the equilibrium distribution of the CV) is maintained, and $\maps$ transforms only the degrees of freedom orthogonal to $\s$, which are distributed according to $p_B(\x|\s)$.
Similarly to TFEP, to maximize the overlap and achieve instantaneous convergence, the optimal map $\maps^*$ transforms $p_B(\x|\s) \to p_A(\x|\s)$ for all values of $\s$, which implies $w_{AB'}(\x|\maps^*) = \Delta f_{AB}(\s)$.

The derivation and characterization of the learning problem for $\maps^*$ follow arguments similar to those used above for TFEP and it is detailed in the Supplemental Material~\cite{supplementalmaterial}. We discuss here the main results. 
The first is that  $\maps^*$ can be learned by minimizing the same log-likelihood in Eq.~(\ref{eq:loss-tfep}) with the key difference that the samples in the training dataset can be obtained from simulations employing an arbitrary biasing potential of the form $V(\s)$.
This has two crucial consequences:
(i) it enables the use of CV-based enhanced sampling techniques in the reference simulation;
(ii) it allows allocating more data points in the training dataset to areas of the CV that would be otherwise poorly represented like, for instance, transition states.
In our experience, both are critical to obtain an improved free energy surface across the whole range of $\s$.

The second is that the following estimator for $\Delta f_{AB}(\s)$ can be derived for reference simulations performed with umbrella sampling and metadynamics
\begin{equation}
    \Delta \hat{f}_{AB}(\s, D | \maps) = -\log \frac{\sum_i \alpha_i e^{-w_{AB'}(\x_i|\maps)} \deltas{\x_i}}{\sum_i \alpha_i \deltas{\x_i}}
\label{eq:dfs-estimator-bias}
\end{equation}
where the weights $\alpha_i$ take different values depending on whether US, MetaD, or no biasing potential was used.
The third and final result is that the maximum-likelihood map is subject to systematic error when $\Delta \hat{f}_{AB}$ is evaluated on the training dataset according to
\begin{equation}
    \sum_s \frac{N_s}{N_{\train}}  \Delta f_{AB}(\s) \ge \sum_s \frac{N_s}{N_{\train}} \Delta \hat{f}_{AB}(\s, D_{\train} | \bthetaml) \; ,
\label{eq:train-dfes-bound}
\end{equation}
where $N_s = \sum_i^{N_{\train}} \deltas{\x_i}$ is the number of samples in bin $\s$, and the summation goes over all bins.
Note that the implications of Eq.~(\ref{eq:train-dfes-bound}) are slightly different from TFEP (see Eq.~(\ref{eq:train-df-bound})).
In particular, Eq.~(\ref{eq:train-dfes-bound}) states that overfitting causes the free energy surface to be underestimated on $D_{\train}$ in an average sense.
As a result, fortuitous cancellation/amplification of error may arise when the FES thus obtained is used to compute free energy differences along $\s$.

We now test the methodology by investigating a simple reaction, the $\mathrm{S_N2}$ reaction \ce{CH3F + Cl- \to CH3Cl + F- } in vacuo.
We evaluated the FES of the reaction at a relatively high level of theory (MP2) from reference data based on semi-empirical PM6 calculations~\cite{stewart2007optimization}.
Four independent 100~ns simulations were performed with well-tempered MetaD~\cite{laio2002escaping,barducci2008well}.
As with any CV-based methodology, the choice of the collective variable is critical.
Here, we use a linear combination of the $\mathrm{C}{\text -}\mathrm{F}$ and $\mathrm{C}{\text -}\mathrm{Cl}$ distances, which was shown to describe well this reaction~\cite{piccini2019accurate}.

\begin{figure}[t]
    \includegraphics{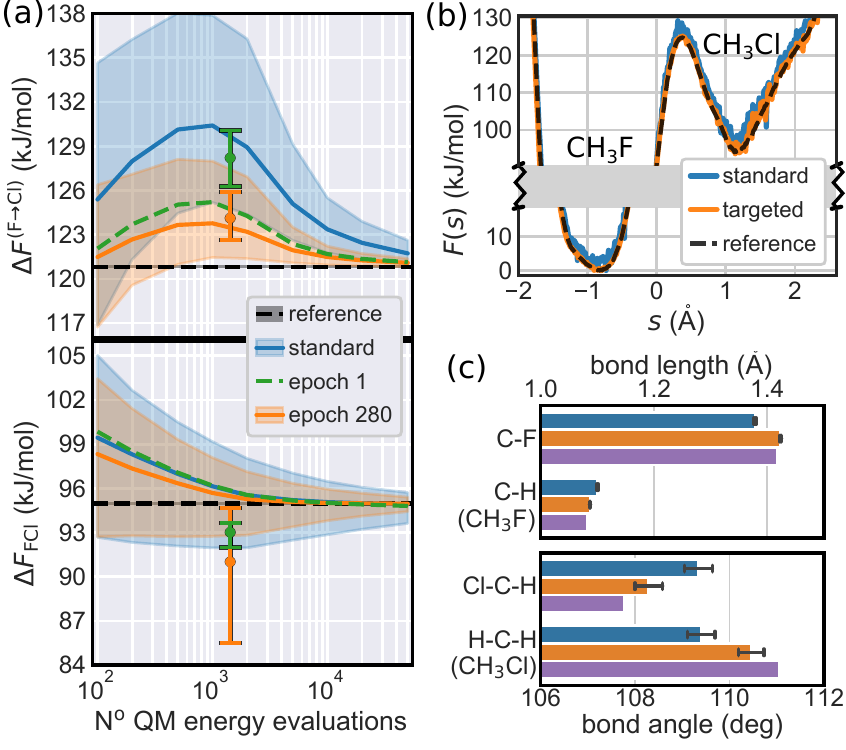}
    \caption{\label{fig:sn2-reaction} (a) Free energy barrier (top) and free energy difference (bottom) for the $\mathrm{S_N2}$ reaction as a function of the number of MP2 energy evaluations computed with standard (blue) and targeted reweighting after 1 (green) and 280 (orange) epochs of training.
    Lines and shaded areas represent means and 95\% confidence intervals obtained by bootstrapping over the evaluation set.
    Dots and error bars were instead computed on 50 different training sets.
    (b) Free energy profile along $s$.
    (c) Comparison of the average bond lengths (top) and angles (bottom) between the simulated (blue), mapped (orange), and MP2-optimized geometries (purple).}
\end{figure}

Fig.~\ref{fig:sn2-reaction}(a) shows the convergence speed of standard and targeted FEP in computing the free energy barrier $\Delta f^{(\mathrm{F} \to \mathrm{Cl})}$~\cite{vanden2005transition,bal2020free} and the free energy difference between products and reactants $\Delta f_{\mathrm{ClF}}$.
Thus,  the precision of the estimates can visibly improve on the evaluation dataset even after a single epoch of training in low-data regimes ($N_{\train} = 1428)$.
The learning efficiency is largely a consequence of the reference potential already providing a very good approximation to the target.
Therefore, if the parameters of the NN are initialized so that the map equals the identity function, the training starts already close to a good solution.
On the other hand, rapidly approaching the maximum-likelihood map also means that the effect of overfitting is apparent very soon, and indeed the systematic error on the training set visibly exceeded that on $D_{\eval}$ after only a single epoch of training.

Due to the asymmetric free energy profile, only approximately 1/5 of the configurations represented the \ce{CH3Cl} state in the training and evaluation datasets (see Fig.~\ref{fig:sn2-reaction}(b)).
In addition, we observed that the fluoride ion energetic interactions with the \ce{CH3Cl} hydrogens were much stronger when using PM6 than with MP2, which decreased the overlap between the two Hamiltonians.
As a result, the FES converged more rapidly for the \ce{CH3F} state than for \ce{CH3Cl}.
To prioritize instead the analysis of high free energy states, one can change the biasing potential employed in the simulation or subsample the trajectory to allocate more data points to the desired areas of the FES.

To verify that the learned map did capture the physics of the target Hamiltonian, we compared the mapped configurations to those obtained by performing geometry optimization at the MP2 level of theory.
We found that only a few average bond lengths and angles (shown in Fig.~\ref{fig:sn2-reaction}(c)) were changed by the mapping with any statistical significance, and in all cases, the NN pushed the average structures closer to the MP2-optimized ones.
While we are aware that the anharmonic thermal fluctuations of bond lengths and angles sampled during the simulation prevent an exact comparison to the MP2-optimized structures, the trend is quite apparent.

\textit{Outlook.}---In this Letter, we showed how NN-based mapping functions can be used to improve the accuracy and convergence of free energy differences and FES estimates starting from a (possibly biased) reference potential.
A key contribution is the characterization of the systematic error caused by overfitting, which effectively enables the application of TFEP theory with maps trained solely from the reference distribution.
The work paves the way for further applications of this methodology to different problems (e.g., binding, enzymatic reactions) with any pair of Hamiltonians (e.g., force fields and NN-based potentials).
Because of the expressivity of normalizing flows, which are capable of learning high-dimensional maps between very different distributions~\cite{noe2019boltzmann}, we expect the method to scale well to systems with larger number of atoms such as proteins.
Moreover, the method could in principle be exploited to study transition state geometries at higher levels of theory.

A challenge in computing the FES with this methodology is the implementation of the CV-preserving condition in Eq.~(\ref{eq:cv-preserving-condition}).
While this is relatively easy to enforce in the NN architecture for simple geometric CVs commonly used, for example, in (bio)chemical reactions~\cite{thirman2021elusive,ludwig2020subsurface,pan2019accelerated}, it is not obvious how to realize it efficiently with highly nonlinear variables such as those based on neural networks~\cite{bonati2020data,ravindra2020automatic,hernandez2018variational,wehmeyer2018time}.
Further work will be needed to investigate how this condition can be relaxed or approximated to extend the applicability of the method to such cases. \newline

\begin{acknowledgments}
AR would like to thank GiovanniMaria Piccini and Emiliano Ippoliti for valuable advice concerning the setup of the semi-empirical and QM calculations.
The authors gratefully acknowledge the computing time granted through JARA on the supercomputer JURECA-DC~\cite{krause2018jureca} at Forschungszentrum J{\"u}lich (Project ID: trp2020) and the computational resources provided by RWTH Aachen University.
The project received funding from the Helmholtz European Partnering program ("Innovative high-performance computing approaches for molecular neuromedicine").
PC acknowledges financial support from Deutsche Forschungsgemeinschaft via the Research Unit FOR2518 "Functional Dynamics of Ion Channels and Transporters -- DynIon", project P6.
PC also acknowledges the Human Brain Project funded by the European Union’s Horizon 2020 Framework Programme for Research and Innovation under the Specific Grant Agreement No. 945539 (Human Brain Project SGA3).
\end{acknowledgments}


\providecommand{\noopsort}[1]{}\providecommand{\singleletter}[1]{#1}%
%

\newpage

\setcounter{equation}{0}
\renewcommand{\theequation}{S\arabic{equation}}
\renewcommand\thepage{S\arabic{page}}
\setcounter{page}{1}

\section{Supplemental Material}

\section{CALCULATION OF THE FES WITH TARGETED METHODS}

In this section, we first provide a detailed derivation of the exact identity that can be used to compute the perturbation term $\Delta f_{AB}(\s)$ with targeted methods.
Then, because in calculations of the FES the CV is typically accelerated using biasing potentials, $\Delta f_{AB}(\s)$ must be estimated using reweighting techniques.
We thus provide targeted estimators for two common enhanced sampling methods, namely umbrella sampling and metadynamics.

\subsection{Unbiased simulation}

An invertible map $\maps: \Gamma_A \to \Gamma_B$ satisfying the condition
\begin{equation}
    \s\left(\maps(\x)\right) = \s(\x)
\label{eq:sm-cv-preserving-condition}
\end{equation}
transforms a distribution $B$ to a distribution $B'$ through a change of variable such that the configurational partition function restricted to $\s$ is given by
\begin{equation}
\begin{split}
    Z_{B'}(\s) &= \int_{\Gamma_A} e^{-u_{B'}(\x|\maps)} \deltas{\x} \dd\x \\
    &= \int_{\Gamma_A} e^{-u_{B}(\maps(\x)) + \logJm[\maps]{\x}} \deltas{\x} \dd\x \\
    &= \int_{\Gamma_B} e^{-u_B(\y)} \deltas{\y} \dd\y = Z_B(\s) \; .
\end{split}
\end{equation}
Thus, the partition function $Z_B(\s)$ is preserved under the transformation.
As a consequence, $f_B(\s) = - \log Z_B(\s) = f_{B'}(\s)$ and we can write
\begin{equation}
\begin{split}
    \Delta f_{AB}(\s) &= \frac{\int_{\Gamma_A} e^{-u_{B'}(\x|\maps)} \deltas{\x} \dd\x}{Z_A(\s)} \\
    &= \frac{\int_{\Gamma_A} e^{-u_A(\x) - w_{AB'}(\x|\maps)} \deltas{\x} \dd\x}{Z_A(\s)} \\
    &= \int_{\Gamma_A} p_A(\x|\s) e^{-w_{AB'}(\x|\maps)} \dd \x \\
    &= \frac{\langle e^{-w_{AB'}(\x|\maps)} \deltas{\x} \rangle_A}{p_A(\s)}\; ,
\end{split}
\label{eq:sm-dfs-identity}
\end{equation}
where $p_A(\s) = Z_A(\s) / Z_A$.
The perturbation term can be estimated from a dataset $D = \{ \x_i \}$ of $N$ samples from an unbiased molecular simulation with
\begin{equation}
    \Delta \hat{f}_{AB}(\s, D|\maps) = \frac{1}{N_s} \sum_i^N \deltas{\x_i} e^{-w_{AB'}(\x|\maps)} \; ,
\label{eq:sm-dfs-estimator-unbiased}
\end{equation}
where $N_s = \sum_i^N \deltas{\x_i}$ is the number of samples in bin $\s$.

\subsection{Umbrella sampling}

In umbrella sampling, one performs $K$ independent simulations, each using a different biasing potential $V_k(\s)$.
Let $N_k$ be the number of samples for the $k$-th window.
We can aggregate all samples from all windows into a single dataset of size $N = \sum_k N_k$.
Then, the aggregate data can be thought as sampled from the mixture distribution~\cite{shirts2017reweighting}
\begin{equation}
    p_U(\x) = \sum_k^K \frac{N_k}{N} p_k(\x) = \sum_k^K \frac{N_k}{N} e^{f_k - u_A(\x) - \beta V_k(\s(\x))} \; ,
\label{eq:sm-umbrella-sampling-distribution}
\end{equation}
where $p_k(\x)$ is the equilibrium distribution of the $k$-th window.
Following the derivation for the standard estimator~\cite{li2018accelerated} (i.e., not targeted), the free energies $f_k$ (defined up to an arbitrary constant) can be computed using the multistate Bennett acceptance ratio (MBAR)~\cite{shirts2008statistically,shirts2017reweighting}.
Then, let us define the weight of configuration $\x_i$ as
\begin{equation}
    \alpha_i = \frac{p_A(\x_i)}{p_U(\x_i)} = \frac{N e^{f_A}}{\sum_k^K N_k e^{f_k - \beta V_k(\s(\x_i))}} \; .
\label{eq:sm-umbrella-sampling-weight}
\end{equation}
The perturbation term can be computed as
\begin{equation}
    \Delta \hat{f}_{AB}(\s, D|\maps) = - \log \frac{\sum_i^N \alpha_i \deltas{\x_i} e^{-w_{AB'}(\x|\maps)}}{\sum_i^N \alpha_i \deltas{\x_i}} \; .
\label{eq:sm-dfs-estimator-bias}
\end{equation}
Note that the constant term $e^{f_A}$ appearing in Eq.~(\ref{eq:sm-umbrella-sampling-weight}) always simplifies in Eq.~(\ref{eq:sm-dfs-estimator-bias}) and is not required for the calculation.
Moreover, in the limiting case of zero-width bins, Eq.~(\ref{eq:sm-dfs-estimator-bias}) reduces to Eq.~(\ref{eq:sm-dfs-estimator-unbiased}).

\subsection{Metadynamics}

We focus here on well-tempered metadynamics (WTMetaD), in which the biasing potential $V(\s, t)$ depends on the time $t$.
After discarding an initial transient, the distribution sampled with WTMetaD can be written as~\cite{bonomi2009reconstructing,tiwary2015time}
\begin{equation}
    p_M(\x, t) = e^{-\beta \left[V(\s(\x), t) - c(t) \right]} p_A(\x) \; ,
\label{eq:sm-quasi-static-distribution}
\end{equation}
where $c(t)$ is the total reversible work done by the biasing potential at time $t$.
Thus, if we define the weight as
\begin{equation}
    \alpha_i = e^{\beta \left[ V(\s(\x_i), t_i) - c(t_i) \right]}
\label{eq:sm-metadynamics-weight}
\end{equation}
the perturbation term is given once again by Eq.~(\ref{eq:sm-dfs-estimator-bias}).
The estimator generalizes that in Ref.~\cite{piccini2019accurate}, which is recovered by setting $\maps$ to the identity function.
Moreover, the estimator reduces to Eq.~(\ref{eq:sm-dfs-estimator-unbiased}) in the limit of zero-width bins and $t \to \infty$, when the biasing potential is fully converged and the dependency on $t$ can be ignored.

\section{FORMULATION OF THE LEARNING PROBLEM FOR $\maps^*$}

Here, we show that the optimal map $\maps^*$ for Eq.~(\ref{eq:sm-dfs-identity}) can be obtained by minimizing the same negative log-likelihood $\mathcal{L}(\btheta)$ derived for TFEP and using a training dataset $D_{\train} = \{ \x_i, t_i \}$ collected by running an enhanced sampling simulation.
As discussed in the main text, we aim to learn the optimal map $\maps^*$ that induce a distribution $B'$ so that
\begin{equation}
    p_{B'}(\x|\s, \maps^*) = p_A(\x|\s)
\label{eq:sm-optimal-maps-defining-condition}
\end{equation}
for all $\s$.
Let us rewrite the sampled distribution in the form
\begin{equation}
    p_V(\x, t) = p_A(\x|\s(\x)) \: p_V(\s(\x), t) \; .
\end{equation}
The expression of $p_V(\s, t)$ depends on the reference simulation.
In the case of unbiased simulations, $p_V(\s, t) = p_A(\s)$.
For umbrella sampling it is given by
\begin{equation}
    p_V(\s, t) = Z_A(\s) \sum_k^K \frac{N_k}{N} e^{f_k - \beta V_k(\s)}  \; .
\end{equation}
Finally, for metadynamics we have
\begin{equation}
    p_V(\s, t) = p_A(\s) e^{-\beta \left[ V(\s,t) - c(t)\right]} \; .
\end{equation}
The key observation is that in all three cases $p_V(\s, t)$ depends on $\x$ only through $\s(\x)$.
Thus, due to Eq.~(\ref{eq:sm-cv-preserving-condition}), it does not depend on the parameters $\btheta$, and using Eq.~(\ref{eq:sm-optimal-maps-defining-condition}), the likelihood of observing $N_{\train}$ independent samples from the sampled distribution can be written as
\begin{equation}
\begin{split}
    p_V(D_{\train}) &= \prod_i^{N_{\train}} p_A(\x_i|\s(\x_i)) \: p_V(\s(\x_i), t_i)
    \\
    &\propto \prod_i^{N_{\train}} p_{B'}(\x_i|\s(\x_i), \maps^*)
    \\
    &= \prod_i^{N_{\train}} e^{-u_{B}(\maps^*(\x_i)) + \logJm[\maps^*]{\x_i}} \; .
\end{split}
\label{eq:sm-dfs-likelihood}
\end{equation}
After multiplying by $\prod_i e^{u_A(\x_i)}$, we obtain once again the negative log-likelihood
\begin{equation}
    \mathcal{L}(\btheta) = \frac{1}{N_{\train}} \sum_i^{N_{\train}} w_{AB'}(x_i | \btheta) \; .
\label{eq:sm-loss-tfep}
\end{equation}

Fundamentally, this property is a consequence of the fact that the optimal map $\maps^*$ is invariant to changes in the equilibrium distribution of the CV.
In other words, $\maps^*$ transforms $p_B(\x|\s)p(\s) \to p_A(\x|\s)p(\s)$ for any choice of $p(\s)$.
We stress, however, that while the minimum of Eq.~(\ref{eq:sm-loss-tfep}) does not change with $p(\s)$, the intermediate solutions explored during the minimization do.
In particular, this means that one is free arbitrarily allocate more data points to areas of the CV that would be otherwise poorly represented in the training set and thus learned inefficiently.

\subsection{Asymptotic behavior of $\mathcal{L}(\btheta)$}

In the limit of infinite sampling (i.e., $N_{\train} \to \infty$), the WTMetaD biasing potential converges, and we can ignore the time dependence of $p_V$ without loss of generality.
Define the distribution
\begin{equation}
    p_{V'}(\x) = p_{B'}(\x|\s(\x), \maps) \; p_V(\s(\x)) \; .
\end{equation}
Then, minimizing $\mathcal{L}(\btheta)$ is equivalent in the limit of $N_{\train} \to \infty$ to minimize the following KL divergence
\begin{equation}
\begin{split}
    D_{\mathrm{KL}}\left[ p_V || p_{V'} \right] &= \int_{\Gamma_A} p_V(\x) \log \frac{p_V(\x)}{p_{V'}(\x)} \dd\x \\
    &= \int_{\Gamma_A} p_V(\x) \log \frac{p_A(\x|\s(\x))}{p_{B'}(\x|\s(\x), \maps)} \dd\x \\
    &= \langle w_{AB'}(\x, \maps) \rangle_V - \langle \Delta f_{AB}(\s(\x)) \rangle_V
\end{split}
\label{eq:sm-kl-div-dfs}
\end{equation}
because the second term in the last line does not depend on the map.
The KL divergence in Eq.~(\ref{eq:sm-kl-div-dfs}) is minimized if and only if $p_{B'}(\x|\s,\maps) = p_A(\x|\s)$ for all values of $\x$ and $\s$.
Thus, in the limit of $N_{\train} \to \infty$, minimizing $\mathcal{L}(\btheta)$ yields the correct optimal map $\maps^*$.

\subsection{Systematic error on the training dataset}

Similarly to TFEP, with a finite training dataset we have, by definition
\begin{equation}
    \frac{1}{N_{\train}} \sum_i^{N_{\train}} w_{AB'}(x_i | \bthetaoptim) \ge \frac{1}{N_{\train}} \sum_i^{N_{\train}} w_{AB'}(x_i | \bthetaml) \; .
\label{eq:sm-thetaoptim-vs-thetaml}
\end{equation}
Using Jensen's inequality and that $w_{AB'}(\x|\maps^*) = \Delta f(\s(\x))$, we obtain
\begin{equation}
    \sum_s \frac{N_s}{N_{\train}}  \Delta f_{AB}(\s) \ge \sum_s \frac{N_s}{N_{\train}} \Delta \hat{f}_{AB}(\s, D | \bthetaml) \; ,
\label{eq:sm-train-dfes-bound}
\end{equation}
where $N_s$ and $\Delta \hat{f}_{AB}$ are defined as in Eq.~(\ref{eq:sm-dfs-estimator-unbiased}).

\section{TARGETED ENSEMBLE AVERAGES OF AN OBSERVABLE}

In the analysis of the $\mathrm{S_N2}$ reaction, free barriers were computed following Refs.~\cite{vanden2005transition,bal2020free} with
\begin{equation}
    \Delta f_B^{(1 \to 2)} = f_B(\s_{\mathrm{TS}}) - f_B^{(1)} - \log \lambda \langle | D(\x) | \rangle_{B|\s} \; ,
\label{eq:sm-reaction-rate}
\end{equation}
where $\s_{\mathrm{TS}}$ is the value of the CV defining the transition state, $f_B^{(1)}$ is the free energy of the first state (which can be obtained from $f_B(\s)$ by integration), and $\lambda$ is a volume-scaling term forcing the argument of the logarithm to be dimensionless.
In particular, the last term in Eq.~(\ref{eq:sm-reaction-rate}) corrects for the dependency on the parametrization chosen for the CV and requires the calculation of an ensemble average of the determinant of the matrix $D(\x) = J_s(\x) \cdot J_s^T(\x)$, where $(J_s)_{ij} = \frac{\partial s_i}{\partial x_j}$ is the Jacobian matrix of $\s(\x)$~\cite{bal2020free}.

More generally, the problem requires the calculation of the ensemble average of an observable $O(\x)$ restricted to $\s$, which we can write as
\begin{equation}
\begin{split}
    \langle O(\x) \rangle_{B|\s} &= \int_{\Gamma_B} O(\y) p_B(\y|\s) \dd\y \\
    &= \int_{\Gamma_A} O(\maps(\x)) p_{B'}(\x|\s,\maps) \dd\x \\
    &= \int_{\Gamma_A} O(\maps(\x)) \frac{p_{B'}(\x|\maps)}{p_B(\s)} \deltas{\x} \dd\x \\
    &= \frac{ \langle O(\maps(\x)) e^{- w_{AB'}(\x|\maps)} \deltas{\x} \rangle_A}{e^{-\Delta f_{AB}(\s)} p_A(\s)} \; ,
\end{split}
\label{eq:sm-ensemble-average-identity-s}
\end{equation}
where in the last line we multiplied and divided by $p_A(\x) p_A(\s)$.
In the case of a biased reference simulation, an asymptotically unbiased estimator for Eq.~\ref{eq:sm-ensemble-average-identity-s} is given by
\begin{equation}
    \overline{O(\s)} = \frac{\sum_i^N O(\maps(\x_i)) \alpha_i \beta_i \deltas{\x_i}}{\sum_i^N \alpha_i \beta_i \deltas{\x_i}} \; ,
\label{eq:sm-estimator-observable-s}
\end{equation}
where the weights $\alpha_i$ are given by Eq.~(\ref{eq:sm-umbrella-sampling-weight}) or Eq.~(\ref{eq:sm-metadynamics-weight}), depending on whether umbrella sampling or metadynamics is used, and $\beta_i = e^{-w_{AB'}(\x_i|\maps)}$.
Eq.~(\ref{eq:sm-estimator-observable-s}) is easily obtained by plugging in Eq.~(\ref{eq:sm-ensemble-average-identity-s}) the expression of the targeted estimator $\Delta \hat{f}_{AB}$ (see Eq.~\ref{eq:sm-dfs-estimator-bias}) for $\Delta f_{AB}(\s)$.
Its main advantage is that one can reuse the values of $w_{AB'}(\x_i|\maps)$ computed for the free energy difference, thus avoiding further need of single point energy calculations at the target level of theory.

Similarly, the global ensemble average of $O(\x)$ can be written as
\begin{equation}
\begin{split}
    \langle O(\x) \rangle_{B} &= \int_{\Gamma_B} O(\y) p_B(\y) \dd\y \\
    &= \int_{\Gamma_A} O(\map(\x)) p_{B'}(\x|\map) \dd\x \\
    &= \int_{\Gamma_A} O(\map(\x)) \frac{p_{B'}(\x|\map)}{p_A(\x)} \dd\x \\
    &= \langle O(\map(\x)) e^{\Delta f_{AB} - w_{AB'}(\x|\map)} \rangle_A \; ,
\end{split}
\end{equation}
and can be estimated with
\begin{equation}
    \overline{O} = \frac{\sum_i^N O(\map(\x_i)) \alpha_i \beta_i}{\sum_i^N \alpha_i \beta_i} \; ,
\label{eq:sm-estimator-observable-global}
\end{equation}
Note that the map in Eq.~(\ref{eq:sm-estimator-observable-global}) does not necessarily have to satisfy the CV-preserving condition in Eq.~(\ref{eq:sm-cv-preserving-condition}), while the map in Eq.~(\ref{eq:sm-estimator-observable-s}) does.

Finally, note that the choice of the map affects the variance of the estimate.
In particular, in the case of an unbiased reference simulation, when the optimal TFEP map $\map^*$ is chosen $\beta_i = e^{-\Delta f_{AB}}$, and we find
\begin{equation}
\begin{split}
    \mathrm{Var}_A\left[ \; \overline{O} \; \right] &= \frac{\mathrm{Var}_A\left[ \; O(\map^*(\x)) \; \right]}{N} = \frac{\mathrm{Var}_B\left[ O(\x) \right]}{N} \; ,
\end{split}
\label{eq:sm-var-estimator-observable-global}
\end{equation}
where we used that performing a change of variable with the inverse optimal map $(\map^*)^{-1}(\x)$ transforms $p_A(\x) \to p_B(\x)$.
Eq.~(\ref{eq:sm-var-estimator-observable-global}) states that the precision of the estimator would be identical as if we were sampling directly from the target distribution $B$.
A similar conclusion can be drawn for the variance of $\overline{O(\s)}$ and the optimal FES map $\maps^*$.

From the variance reduction methods literature, it is known that this is not the minimum theoretical variance achievable for the estimate (see for example Ref.~\cite{hesterberg1988advances} for the theoretical limit on importance sampling estimators).
However, in the context of atomistic simulations, we expect this will likely result in a reduced statistical error.

\section{NUMERICAL EXPERIMENTS}

The code and the simulation input files used in this work can be found at Ref.~\cite{code}.

\subsection{Double-well potential}

The reference and target distributions were modeled with a mixture of two Gaussians with different mean and variance matrix so as to allow independent sampling.
The potential energy of the two systems was then defined as $u_A(\x) = -\log p_A(\x) - \Delta f_{AB}$ for the reference and  $u_B(\x) = -\log p_B(\x)$ for the target, where $\Delta f_{AB}$ was the arbitrary exact value of the free energy diafference between the two distributions.
The inverse autoregressive flow (IAF) was implemented with PyTorch~1.7~\cite{paszke2019pytorch}, and it was composed by four layers.
Each layer used a 2-layer MADE~\cite{germain2015made} network for the conditioner and an affine transformer~\cite{papamakarios2021normalizing}.
The order of the variables was reversed in each layer.
For each tested training dataset size, the NN was trained on 40 different randomly-generated datasets for 10000 epochs using the ADAM optimizer~\cite{kingma2014adam} with default parameters and a learning rate of 0.001.
Each network was then tested on 50 independent evaluation sets of size 100.

\subsection{$\mathrm{S_N2}$ reaction}

To collect the reference data, we used the semi-empirical PM6 potential as implemented in the AMBER~16~\cite{case2016amber16} and we performed four independent 105~ns WTMetaD simulation with PLUMED~2.6~\cite{tribello2014plumed}.
The first 5~ns was discarded so that only the quasi-static portion of the simulation was analyzed.
We used a SCF convergence criteria of $10^{-8}$~kcal/mol, a timestep of 0.5~fs and a Langevin thermostat to control the temperature at 300~k.
The system was confined by applying an harmonic potential wall when either of the two distances $d_{\mathrm{C}{\text -}\mathrm{F}}$ or $d_{\mathrm{C}{\text -}\mathrm{Cl}}$ exceeded 4~\r{A}.
The WTMetaD was implemented by depositing every 50 steps a Gaussian hill with height 2~kJ/mol, sigma 1~\r{A} and bias factor 50.
The MP2 energy evaluations and gradients required for the training were performed with Psi4~1.3~\cite{parrish2017psi4} using the aug-cc-pVDZ basis set with frozen core orbitals, the density fitting approximation, and an SCF energy convergence criteria of $10^-8$~Hartree.
We used the following CV, which was developed in a previous work for this reaction~\cite{piccini2019accurate}
\begin{equation}
    s = \alpha d_{\mathrm{C}{\text -}\mathrm{F}} + \beta d_{\mathrm{C}{\text -}\mathrm{Cl}} \; ,
\label{eq:sm-sn2-reaction-cv}
\end{equation}
where $d_{\mathrm{C}{\text -}\mathrm{F}}$ and $d_{\mathrm{C}{\text -}\mathrm{Cl}}$ are the distances between the carbon and the fluorine and chlorine atoms, respectively, and $\alpha = 0.81$ and $\beta = -0.59$ are constant coefficients that were determined using the HLDA method~\cite{mendels2018collective,piccini2018metadynamics}

A reference value for the FES and the free energy difference was computed with standard FEP on all the samples after aggregating the 4 independent simulations for a total of $8 \cdot 10^6$ MP2 point energy evaluations.
The main training dataset was obtained by subsampling the trajectory with a constant time interval of 70~ps.
Furthermore, 50 more training datasets were generated by randomly subsampling the trajectory to evaluate the systematic error introduced by overfitting.

To improve the speed of the learning, we removed the translational and rotational symmetries of the system by centering the carbon atom in the origin and placing the chlorine on the z-axis and an hydrogen atom on the xz-plane.
Moreover, to enforce Eq.~(\ref{eq:sm-cv-preserving-condition}), we implemented the mapping function using a conditional Masked Autoregressive Flow (MAF) architecture~\cite{papamakarios2017masked} in which the value of $s$ was not altered but still included in the input to affect the mapping of its orthogonal degrees of freedom.
In particular, the only degree of freedom of the Cl (the z coordinate) was not mapped by the NN, and its value was fixed after the mapping in order to preserve the CV.
In practice, only 11 degrees of freedom of the 6 atoms were mapped by the NN.
The MAF was composed by 12 layers, each using a 2-layer MADE for the conditioner and an affine transformer and inverting the order of the variable at each layer.
The NN was optimized for 280 epochs with batch size 256 using the ADAM optimizer and a cyclical learning rate going from 0.0001 to 0.001 and back in 10 epochs.

The network was applied to the training datasets and 10000 evaluation sets of increasing size by randomly subsampling the trajectory to perform bootstrap analysis.
The free energy barriers were computed using Eq.~(\ref{eq:sm-reaction-rate}) and the estimator in Eq.~(\ref{eq:sm-estimator-observable-s}).

Finally, to perform the analysis of the mapped geometries, we randoml selected 512 configurations from each of the two metastable states and optimize them using Psi4.
The \ce{CH3F} state was defined to have the CV within the interval $(-1.5, -0.5)$~{\AA}, and the \ce{CH3Cl} state within $(0,8, 1.5)$~{\AA}.
Less than 10 optimizations in each state failed to converge after 2000 iterations and where ignored.
As the reference value for the optimized geometry, we took only the optimized configuration with the minimum energy in each basin.
We compared to these configurations those sampled with PM6 before and after mapping with the trained network.
The configurations in each state were selected after subsampling the trajectory with a 1~ps time interval, and they where selected to have the same value of the CV as the optimum geometry $\pm 0.01$~{\AA}.
All the combinations were considered (and averaged) when computing distances and angles relative to hydrogen atoms.

\end{document}